\documentclass[twocolumn,showpacs,floatfix]{revtex4}
\usepackage{dcolumn,amssymb}
\usepackage{amsmath,epsfig}
\usepackage{rotating}
\newcommand{\C}{\mathbb{C}}

\newcommand{\R}{\mathbb{R}}

\newcommand{\cM}{{\cal M}} 
\newcommand{\cO}{{\cal O}}

\newcommand{\half}{\frac{1}{2}}
\newcommand{\bea}{\begin{eqnarray}}
\newcommand{\eea}{\end{eqnarray}}
\def\be#1\ee{\begin{equation}#1\end{equation}}

\newcommand{\bra}{\langle}
\newcommand{\ket}{\rangle}

\def\be#1\ee{\begin{equation}#1\end{equation}}

\begin{document}

\title
{The Complex Langevin method: When can it be trusted?}

\author {Gert Aarts} \email{g.aarts@swan.ac.uk} 
\affiliation {Department of Physics, Swansea University, 
Swansea, United Kingdom}
\author{ Erhard Seiler}\email{ehs@mppmu.mpg.de}
\affiliation{Max-Planck-Institut f\"ur Physik
(Werner-Heisenberg-Institut), 
M\"unchen, Germany}
\author{Ion-Olimpiu Stamatescu} \email{I.O.Stamatescu@thphys.uni-heidelberg.de}
\affiliation{ Institut f\"ur Theoretische Physik, Universit\"at Heidelberg
and FEST, 
Heidelberg, Germany}
\date{December 17, 2009}
\begin{abstract}\noindent
We analyze to what extent the complex Langevin method, which is in 
principle capable of solving the so-called sign problems, can be 
considered as reliable. We give a formal derivation of the correctness and 
then point out various mathematical loopholes. The detailed study of some 
simple examples leads to practical suggestions about the application of 
the method.
\end{abstract}
\pacs {11.15.Ha, 11.15Bt \hfill arXiv:0912.3360 [hep-lat]} 
\maketitle

\section{Introduction}
The complex Langevin method solves in principle the sign problems arising 
in simulations of various systems, in particular in QCD with finite 
chemical potential.
 After being proposed in the early 1980s by Klauder \cite{klauder1} and 
Parisi \cite{parisi}, it enjoyed a certain limited popularity 
\cite{Karsch:1985cb,Damgaard:1987rr}, but very 
quickly certain problems were found. The first one was instability of the 
simulations with absence of convergence (runaways), the second one 
convergence to a wrong limit \cite{Ambjorn:1985iw, klauder2, 
linhirsch,Ambjorn:1986fz}. Nevertheless in recent years the method has 
been revived with sometimes impressive success \cite{bergesstam, berges07, 
Berges:2007nr, aartsstam, aarts1, aarts2, guralnik, guralnik2}. In 
particular the use of adaptive stepsize has eliminated the problem of 
runaways 
\cite{Aarts:2009dg}.

But nagging problems remained due to the lack of clear criteria to decide 
when an apparently convergent simulation actually represented the truth. 
This was linked to the lack of a clear mathematical basis for the method, 
that would at the same time also provide criteria for its applicability. 
The purpose of the present paper is to clarify the situation at least to 
some extent. While we are still not able to close certain mathematical 
gaps and reach a complete analytic solution to the problems that have 
plagued the method, we give some strong numerical evidence that the method 
is correct in some cases and also suggest a plausible explanation for the 
failure in other cases; this leads to some pragmatic conclusions 
suggesting how to proceed in practice in a way that promises credible 
results.

The paper is organized as follows. In Sec.~\ref{secII} we give a formal 
justification of the method, highlighting the assumptions 
underlying the derivation. In Sec.~\ref{secIII} three main questions 
raised by the formal arguments are listed. We then focus on one particular 
issue, boundary effects, in Sec.~\ref{secIV}, and present detailed case 
studies in Sec.~\ref{secV}. Tentative conclusions are given in 
Sec.~\ref{secVI}.

\section{Formal derivations}
\label{secII}

For simplicity we concentrate here on models in which the fields take 
values in flat manifolds ${\cal M}=\R^n$ or ${\cal M}=T^n$, where $T^n$ is 
the $n$ dimensional torus $(S^1)^n$ with coordinates $(x_1,\ldots,x_n)$. 
The complications that arise when the fields live in nontrivial manifolds, 
as is of course the case in QCD, have been successfully dealt with in the 
literature (see for instance Ref.~\cite{batrouni} for real, 
Refs.~\cite{gausthaler, gausterer,berges07, aartsstam} for  complex 
Langevin dynamics). But these complications are not really relevant for 
our discussion.

As is well known, the idea is to simulate a complex measure 
$\exp(-S)dx$, with $S$ a holomorphic function on a real manifold $\cM$, 
by setting up a stochastic process on the complexification $\cM_c$ of 
$\cM$, such that the expectation values of {\it entire holomorphic 
observables} $\cO$ in this stochastic process converge to the ones with 
respect to the complex measure $\exp(-S) dx$.

The complex Langevin equation (CLE) on $\cM_c$ is
\be
dz= -\nabla S dt+dw,
\label{cle0}
\ee
where $dw$ denotes the increment of the Wiener process and the equation is 
to be interpreted as a real stochastic process, namely
\begin{align}
\label{cle1}
dx=&K_x dt +dw,\notag\\
dy=&K_y dt,
\end{align}
with
\begin{align}
K_x=& -{\rm Re} \nabla_x S(x+iy), \notag \\
K_y=& -{\rm Im} \nabla_x S(x+iy).
\end{align}
A slight generalization of  Eq.~(\ref{cle1}) that has been considered and will 
play a role in this investigation is
\begin{align}
dx=&K_x dt +\sqrt{N_R}dw_R,\notag\\
dy=&K_y dt +\sqrt{N_I}dw_I,
\label{cle2}
\end{align}
where $dw_R$ and $dw_I$ are independent Wiener processes, $N_I\ge 0$ and 
$N_R=N_I+1$. This is usually  referred to as complex noise. 
The introduction of a nonzero $N_I$ makes it possible to solve 
the Fokker-Planck equation (see below) numerically and also allows a random 
walk discretization of the complex Langevin process:
\begin{align}
{\delta x(t)} = & \pm \omega_x,  \ \ 
  P_{x,\pm} =  \half\left(1\pm \tanh \left[\frac{\omega_x}{2 N_R}K_x \right]\right),  
 \label{e.rws1}\\
 {\delta y(t)} = & \pm \omega_y,  \ \ 
  P_{y,\pm} = \half\left(1\pm \tanh \left[\frac{\omega_y}{2 N_I} K_y\right]\right),  
 \label{e.rws2}\\
 & \omega_x = \sqrt{2N_R\delta t}, \  \  \  \
 \omega_y =\sqrt{2N_I\delta t}, 
\end{align}
where $P_{\pm}$ are the transition probabilities and we have defined the 
steps $\omega$ such as to have the same $\delta t$ in both sub-processes, 
to ensure correct evolution.

By It\^o calculus, if $f$ is a twice differentiable function on 
$\cM_c$ and 
\be
z(t)=x(t)+iy(t)
\ee 
is a solution of the complex Langevin equation (\ref{cle2}), we have 
\be
\label{ito}
\frac{d}{dt}\left\bra f(x(t),y(t))\right\ket = \left\bra L 
f(x(t),y(t))\right\ket,
\ee
where $L$ is the Langevin operator 
\be
\label{eq:LO}
L=\left[N_R\nabla_x+K_x\right] \nabla_x  
+ \left[N_I \nabla_y+K_y\right] \nabla_y,
\ee
 and $\bra f \ket$ denotes the noise average of $f$ corresponding to the 
stochastic process described by Eq.~(\ref{cle2}). In the standard way 
Eq.~(\ref{cle2}) leads to its dual Fokker-Planck equation (FPE) for the 
evolution of the probability density $P(x,y;t)$,
 \be
\label{realFPE}
\frac{\partial}{\partial t} P(x,y;t)= L^T P(x,y;t),
\ee
with 
\be
L^T=\nabla_x\left[N_R\nabla_x-K_x\right]+
\nabla_y\left[N_I\nabla_y- K_y\right].
\ee
$L^T$ is the formal adjoint (transpose) of $L$ with respect to the 
bilinear (not hermitian!) pairing 
\be
\bra f, P\ket= \int f(x,y) P(x,y) dxdy, 
\ee
i.e.,
\be
\bra Lf,P\ket= \bra f,L^T P\ket.
\ee
Note that the FPE has the form of a continuity equation 
\be
\frac{\partial}{\partial t} P(x,y;t)=\nabla_x J_x+\nabla_y J_y,
\ee
where $(J_x,J_y)$ is the probability current in the $2n$ dimensional space 
$\cM_c$, given by
\be
J_x=\left(N_R\nabla_x-K_x\right)P,\; \;\; J_y= 
\left(N_I\nabla_y-K_y\right)P.
\ee
We will also consider the evolution of a complex density $\rho(x)$ on 
$\cM$ under the following complex FPE 
\be  
\label{complexFPE}
\frac{\partial}{\partial t} \rho(x;t)= L_0^T \rho(x;t),
\ee
where now the complex Fokker-Planck operator $L_0^T$ is  
\be
\label{fpc0}
L_0^T =  \nabla_x \left[\nabla_x+(\nabla_x S(x))\right].
\ee
A slight generalization will be useful: For any $y_0\in \cM$ we consider
a complex Fokker-Planck operator $ L_{y_0}^T$ given by
\be
\label{fpc1}
L_{y_0}^T=\nabla_x \left[\nabla_x+(\nabla_x S(x+iy_0))\right].
\ee
$L_{y_0}^T$ is the formal adjoint of
\be
L_{y_0}=  \left[\nabla_x-( \nabla_x S(x+iy_0))\right]\nabla_x.
\ee
The operators $L_{y_0}^T$ act on suitable complex valued distributions 
(measures) on $\cM$, parameterized by the real variables 
$(x_1,\cdots,x_n)$.  But they do not allow a probabilistic 
interpretation, because they do not preserve positivity.

For any $y_0$ Eq.~(\ref{complexFPE}) with $L_{0}^T$ replaced by 
$L_{y_0}^T$ has the complex density
\be 
\label{rhostat}
\rho_{y_0}(x;\infty)\propto \exp\left[-S(x+iy_0)\right] 
\ee
as its (hopefully unique) stationary solution.

We next consider expectation values. Let $\cO$ be an entire holomorphic 
observable with at most exponential growth; then we set
\be
\label{eq:OP}
\bra \cO\ket_{P(t)}\equiv \frac{\int O(x+iy) P(x,y;t) dxdy}
{\int  P(x,y;t) dxdy}
\ee
and
\be
\bra \cO\ket_{\rho(t)}\equiv \frac{\int O(x) \rho(x;t) dx}
{\int\rho(x;t) dx} .
\ee  
We would like to show that
\be
\bra \cO\ket_{P(t)}=\bra \cO\ket_{\rho(t)},
\ee
provided the initial conditions agree,
\be
\bra \cO\ket_{P(0)}=\bra \cO\ket_{\rho(0)},
\ee
which is true if we choose 
\be 
\label{init}
P(x,y;0)=\rho(x;0)\delta(y-y_0)
\ee 
(for any $y_0$). 
In the limit $t\to\infty$ the dependence on the initial condition should
of course disappear by ergodicity.

The goal is to establish a connection between the `expectation values' 
with respect to $\rho$ and $P$ for the class of observables chosen (entire 
holomorphic with at most exponential growth). The idea is to move the time 
evolution from the densities to the observables and make use of the 
Cauchy-Riemann (CR) equations. Formally (i.e.\ without worrying about 
boundary terms and existence questions) this works as follows: first we
use the fact that we want to apply the complex FP operators $L_{y_0}$
only to functions that have analytic continuations to all of $\cM_c$.
On those analytic continuations we may act with the Langevin operator
\be
\tilde L \equiv \left[\nabla_z-(\nabla_z S(z))\right]  \nabla_z,
\ee
whose action on holomorphic functions agrees with that of $L$, since on 
such functions $\nabla_y=i\nabla_x$ and $\Delta_x =-\Delta_y$ so that the 
difference $L-\tilde L$ vanishes.

The proliferation of Langevin/Fokker-Planck operators may be somewhat 
bewildering, but it is important to realize that $L,\tilde L, L_{y_0}$ are 
really all different operators: while $L$ and $\tilde L$ act on functions 
on $\cM_c$ (i.e.\ functions of $(x_1,y_1,\ldots,x_n,y_n)$), agreeing on 
holomorphic functions, but disagreeing on general functions, $L_{y_0}$ acts 
on functions on $\cM$, i.e.\ functions of $(x_1,\ldots,x_n)$.
  
We now use $\tilde L$ to evolve the observables by the equation
\be
\label{obsevol}
\partial_t \cO(z;t)= \tilde L \cO(z;t)\quad (t\ge 0)
\ee
with the initial condition $\cO(z;0)=\cO(z)$, which is formally solved by
\be
\label{obssol}
\cO(z;t) =  \exp[t \tilde L] \cO(z).
\ee
In Eqs.~(\ref{obsevol}, \ref{obssol}), because of the CR equations, the 
tilde may be dropped, and we will do so now. So we will have  
\be
\label{obsevol2}
\partial_t \cO(z;t)= L \cO(z;t)\quad (t\ge 0),
\ee
with its formal solution
\be
\label{obssol2}
\cO(z;t) = \exp[t L] \cO(z).
\ee
In fact Eq.~(\ref{obsevol}) is also equivalent to the family of equations 
\be
\label{obsevol3}
\partial_t \cO(x+iy_0;t)= L_{y_0} \cO(x+iy_0;t)\quad (t\ge 0)
\quad \forall y_0.
\ee
The first thing to notice is that $\cO(z;t)$ will be holomorphic if
$\cO(z;0)$ 
is. This can be seen as follows: Let $\bar\partial_j$ 
$(i=j,\ldots ,n)$ be the Cauchy-Riemann operators defined by 
\be 
\bar\partial_j\equiv \partial_{x_j}+i\partial_{y_j}.
\ee
Applying this to both sides of Eq.~(\ref{obsevol}) and using the fact 
that by the holomorphy of $S$ $\bar\partial_j$ commutes with $L$, we find
\be
\label{crobsevol}
\partial_t \bar\partial_j \cO(z;t)= 
L \bar\partial_j \cO(z;t)\quad (t\ge 0)\,;
\ee
this is just Eq.~(\ref{obsevol}) again with $\cO(z;t)$ replaced by 
$\bar\partial_j\cO(z;t)$. Under the assumption that $L$ generates a 
semigroup acting on $\bar\partial_j \cO(z;t)$, Eq.~(\ref{crobsevol}) has a 
unique solution; since the initial condition is $\bar\partial_j O(z;0)=0$ 
we conclude that $\bar\partial_j\cO(z,t)=0$ for all $t\ge 0$, i.e.\
$\cO(z;t)$ satisfies the CR equations for all $t\ge 0$ and all 
$j=1,\ldots, n$. So $\cO(z;t)$ is holomorphic in each component $z_j$ 
separately. By Hartogs' theorem \cite{gunningrossi, vladimirov} this 
implies joint holomorphy.

We now consider, for $0\le \tau\le t$, 
\be
F(t,\tau)\equiv \int P(x,y;t-\tau) \cO(x+iy;\tau)dxdy,
\ee
and claim that it  interpolates between the $\rho$ and the $P$ expectations:
\be
F(t,0)= \bra \cO\ket_{P(t)}, \;\;\;\; F(t,t)= \bra \cO \ket_{\rho(t)}.
\ee
The first equality is obvious, while the second one can be seen as 
follows, using Eqs.~(\ref{init}, \ref{obssol2}), 
\begin{align} 
F(t,t)=&\int P(x,y;0) \exp(t L)\cO(x+iy;0)dxdy\notag\\=&
\int \rho(x;0) \left(\exp(tL_{y_0}) \cO(x+iy_0;0)\right)dx\notag\\=&
\int \cO(x+iy_0;0)\left(\exp(tL_{y_0}^T)\rho(x;0)\right)dx\notag\\ = &
\bra \cO \ket_{\rho(t)},
\end{align}
where we only had to assume that we can integrate by parts in $x$ without 
worrying about boundary terms.

Our desired result follows if we can show that $F(t,\tau)$ is independent 
of $\tau$. To see this, we differentiate
\begin{align}
\label{interpol}
\frac{\partial}{\partial \tau} F(t,\tau) = &
-\int \left(L^T P(x,y;t-\tau)\right)\cO(x+iy;\tau)dxdy\notag\\
& + \int  P(x,y;t-\tau) \tilde L \cO(x+iy;\tau) dxdy.
\end{align}
Integration by parts then shows that the two terms cancel, hence 
$\frac{\partial}{\partial \tau} F(t,\tau)=0$  and thus 
\be
\label{equival}
\bra \cO\ket_{P(t)}= \bra \cO \ket_{\rho(t)}. 
\ee
It is important to notice that this holds for all $N_I$; whereas the left 
hand side seems to depend on $N_I$, the right hand side is manifestly 
independent of it. 

If we knew in addition that 
\be
\label{conv}
\lim_{t\to \infty} \bra \cO\ket_{\rho(t)} = \bra \cO \ket_{\rho(\infty)},
\ee
with $\rho(\infty)$ given by Eq.~(\ref{rhostat}) with $y_0=0$, we could 
now conclude that the expectation values of the Langevin process relax to 
the desired values; this convergence would follow if we knew that the
spectrum of $L^T_{y_0}$ lies in a half plane ${\rm Re}\,z\le 0$ and $0$ 
is a nondegenerate eigenvalue. But note that we do not really need 
convergence of $P(x,y;t)$ for Eq.~(\ref{conv}) to hold, since it will 
only be tested against analytic observables $\cO$.

Nevertheless the numerical evidence in many cases points to the existence 
of a unique stationary probability density $P(x,y;\infty)$. The corresponding 
probability currents are divergenceless, but unlike the situation in the 
real Langevin process, they cannot vanish. A general feature of the 
stationary distribution that can be read off the FPE is the following: 
Assume that $(x_0,y_0)$ is a local stationary point of $P$, then
\be
(N_R\Delta_x+N_I\Delta_y)P =
\nabla_x (K_xP)+\nabla_y (K_y P),
\ee
for $x=x_0, y=y_0$.
So if $N_I>0$, a local maximum of $P$ can only occur where the divergence 
of the drift force is negative and a minimum where it is positive. 
For $N_I=0$ the conclusion is even stronger: where the divergence is 
negative (positive), there cannot be a local minimum (maximum) in $x$ for 
fixed $y$. These properties provide some checks on numerical solutions. 

\section{Questions}
\label{secIII}

There are three main questions raised by the formal arguments in the previous 
section:

(1) Can the operators $L, \tilde L, L_{y_0}$ and their transposes be 
exponentiated; in more mathematical language: do these operators generate 
semigroups on some suitable space of functions?

(2) Are the various integrations by parts justified, which underlie the 
shifting of the time evolution from the measure to the observables and back,
or are there boundary terms to worry about?

(3) Are the spectra of $L, L_{y_0}$ and their transposes entirely in the 
left half plane and is 0 a nondegenerate eigenvalue?  

Concerning the first question, there are treatises (see for instance  Refs.\
\cite{daviesbook, nier}) giving rather general sufficient conditions for the 
existence of a semigroup generated by differential operators of the general 
type considered here. Unfortunately it seems that the cases we have to deal 
with here are not covered by those general results; the main difficulties are 
(1) the strong growth of the drift given by the gradients of the action in 
some complex directions and (2) the fact that the drift is not always 
`restoring'.

Question (1) for $L, L^T$ is intimately related to the question 
whether the stochastic process given by the complex Langevin equation 
exists for arbitrary long times. This is not obvious because typically in 
the classical (no noise) limit there are trajectories that go to infinity 
in finite time. While those trajectories occur only for a subset of 
measure zero of initial conditions, it is not obvious what happens after 
adding the noise. On the one hand, the noise will typically kick the 
process away from the unstable trajectories; on the other hand it may also 
kick it near the unstable trajectory, inducing very large excursions of 
the process. We will later illustrate this with some examples.

But let us say that the accumulated numerical evidence points not only to 
the existence of the process for arbitrarily large times, but also to the 
existence of a unique equilibrium measure for the process;  unfortunately 
we could neither find results in the mathematical literature that 
would imply this, nor could we prove ourselves that this is the case.
 
Concerning the exponentiation of $\tilde L, L_{y_0}$, which should be easier, 
we still could not establish mathematically that it is possible on a space 
of functions containing the most obvious observables, such as exponentials.

There is a useful criterion for the existence of a bounded semigroup 
generated by an operator $A$ on a Hilbert space 
\cite{daviesbook}: $A$ generates a bounded semigroup if it is {\it 
dissipative}, i.e. if $A+A^\dagger\le 0$. Unfortunately even in the 
simplest cases of a quadratic $S$ the corresponding Langevin and FP 
operators are {\it not} dissipative. So they can at best generate 
exponentially bounded semigroups; if in addition the spectrum is in the 
left half plane, convergence to the equilibrium should still take place. 

For the second question it would be necessary to have good control over
the falloff of the solutions of the FPE in the imaginary directions: if we 
insert the observables  
$\cO_k(z)\equiv\exp(ikz)$ into Eq.~(\ref{equival}), we get for the Fourier 
transform (Fourier coefficients in the compact case) $\widehat \rho(k;t)$ 
of the complex density $\rho(x;t)$ 
\be 
\widehat \rho(k;t) = \frac{1}{(2\pi)^n}\int P(x,y;t) e^{ikx-ky} dxdy. 
\ee 
This makes sense for all $k$ {\it only} if $P(x,y;t)$ decays more strongly 
than any exponential in imaginary direction. Our case studies described in 
the following sections indicate that this does not seem to be the case: 
in our first example the decay is probably exponential, but not stronger; 
in our second example the decay seems to be even weaker ($O(|y|^{-r})$ 
with $r\approx 2$), so that exponentials cannot be used as observables.
The remainder of the paper is mainly devoted to studying Question (2) in 
some toy models.

Finally let us remark that the third question is more difficult than the 
first one, and again the answer is not known rigorously. But again the 
numerical evidence strongly suggests a positive answer, depending on the
model and the parameter values, in many interesting and relevant 
cases (see e.g.\ Refs.~\cite{aartsstam, aarts2}). 

\section{Boundary effects}
\label{secIV}

In this paper we are mainly concerned with Question (2). Even though we 
have very little analytic control, careful numerical studies reveal that, 
as remarked, the answer is generally {\it no}! In our case studies we 
find indications that the probability density $P(x,y;t)$ indeed relaxes to 
an equilibrium density $P(x,y;\infty)$, but that that limiting density 
decays at best like an exponential for $|y|\to\infty$, in other cases only 
power-like. This limits the class of observables for which the integrations 
by parts can be performed without boundary terms. 

But let us first take a closer look at the integrations by parts 
that occur in the formal arguments of Sec.\ \ref{secII}. The danger lies in a 
possibly insufficient falloff in the imaginary ($y$) directions, whereas 
in the real ($x$) direction we have either compactness or sufficient 
falloff due to the behaviour of the action $S$.

Let us remark that for $N_I>0$ the operators $L$ and $L^T$ are uniformly
strictly elliptic; this is important, because it implies regularity for 
any solution of the stationary FPE \cite{gilbarg}. It is also to be
expected that the semigroup $\exp(tL^T)$ has a smooth (even real analytic)
kernel so that for $t>0$ $P(x,y;t)$ will be smooth (real analytic). This  
is supported by our numerical studies.
 So the problematic point is to show that the two terms on right-hand side of  
Eq.~(\ref{interpol}) actually cancel. This may fail because the observables 
$\cO$ typically grow in imaginary direction, whereas the decay of 
$P(x,y;t)$ (always assuming it exists) may be insufficient to compensate 
for it. 

Let us see in a little more detail how the argument for the independence 
of $N_I$ may fail. For simplicity of presentation we consider the 
one-dimensional case ($n=1$). We write the Langevin operator 
(\ref{eq:LO}) as 
\be
L=L_{N_I=0}+N_I\Delta,
\ee
where $L_{N_I=0}$ is the Langevin operator for $N_I=0, N_R=1$. Then let us
consider
\begin{align}
\label{noisedep}
&\frac{\partial}{\partial N_I} \int P(x,y;t)\cO(x+iy) dxdy \notag \\
& = 
\frac{\partial}{\partial N_I}
\int (e^{t(L_{N_I=0}^T+N_I\Delta)}P(x,y;0))\cO(x+iy) dxdy.
\end{align}
We use the formula
\be
\frac{\partial}{\partial N_I}e^{tL^T}=
\int_0^t d\tau\, e^{\tau L^T}\Delta \,e^{(t-\tau)L^T},
\ee
and integration by parts to rewrite Eq.~(\ref{noisedep}) as
\be
\int_0^t d\tau \int P(x,y;t-\tau) \Delta  \cO(x+iy;\tau) dxdy +X.
\ee
The term denoted by $X$ collects possible boundary terms arising in the
integration by parts. It vanishes only if the decay of $P(x,y;t-\tau)$ is 
strong enough to offset any possible growth of $\cO(x,y;\tau)$; otherwise 
it may either converge to a finite nonzero value or diverge. By the CR 
equations the first term vanishes, but the uncontrolled boundary term $X$ 
remains.

Let us look at a simple example that shows how and when the formal argument
fails.  Consider
\begin{align}
X = \int\left(\Delta P(x,y;t)\right) \cO(x+iy)dxdy\notag\\
-\int P(x,y;t) \Delta \cO(x+iy) dxdy
\end{align}
(where the second term vanishes on account of the CR equations). 
By the formal argument this would be zero, being just a 
boundary term, but careful application of integration by parts, at first 
over the finite domain  $-Y_-<y<Y_+$,  gives
\begin{align}
X=
\lim_{Y_\pm\to\infty}  \int dx \,\Big[ \left(\nabla_y P(x,y;t) \right) 
\cO(x+iy)  & \notag\\
  -  P(x,y;t)\nabla_y \cO(x+iy) \Big] \Big|_{-Y_-}^{Y_+}.  &
\end{align}
Here it is clear that what matters is the combined asymptotic behaviour of 
$P$ and $\cO$, and depending on the observable, $X$ may be zero, finite, or 
divergent. Of course the form of the boundary terms is less simple when 
using integration by parts in Eqs.~(\ref{interpol},\ref{noisedep}), but we 
expect that it is still the decay of the products like $P\cO$, 
$\cO\nabla_y P$, $P\nabla_y\cO$ that is relevant.
   
When trying to investigate the effects of the boundary numerically, one 
would in principle like to use the probability density obtained without a 
cutoff. In practice this is, however, not feasible, and we therefore 
introduce a cutoff in the imaginary direction, which is not sent to 
infinity. Such a device is necessary for the solution of the FPE, and even 
though the CLE does not require it, for the purpose of comparison we also 
introduce it there. This will, however, introduce additional problems with 
the formal arguments relying on the CR equations as well as integration by 
parts.

Concretely we proceed as follows: We restrict each $y_j$ to lie between 
$-Y_-$ and $Y_+$ and impose periodic boundary conditions on both the 
observables and the probability densities. This has a number of 
consequences. Firstly, observables $\cO(x+iy)$ will in general not be 
continuous across the `seam', where we identify $y_j=-Y_-$ with $y_j=Y_+$. 
They can therefore not be interpreted as continuous functions and a priori 
the It\^o formula (\ref{ito}) does not hold (it may still hold in the 
sense of distributions, which should be sufficient for our purposes). 
Furthermore, the jump across the seam will mean that the CR equations are 
no longer satisfied everywhere. For the evolved observables the CR 
equations cannot be expected to hold anywhere exactly, as the violation 
that occurred initially only at the boundary gets propagated everywhere by 
the Langevin evolution. Similarly, the drift in the FPE is expected to be 
discontinuous at the seam. Therefore we have to expect that $P$ has a jump 
there as well; this in turn forces us to interpret the FPE in the sense of 
distributions.

One might wonder whether it would not be better to limit the fluctuations 
in the imaginary direction by introducing a smooth cutoff; but since such 
a smooth cutoff function will necessarily be nonholomorphic it will 
destroy the formal arguments even more;  we therefore stick with the simplest 
choice of a periodic cutoff.

We conclude therefore that the introduction of a cutoff $-Y_-<y_j<Y_+$ and 
imposing periodic boundary conditions leads to a breakdown of the formal 
arguments given in Sec.\ \ref{secII}. Although it is difficult to quantify 
precisely the effect of this, it seems reasonable to expect that it is 
still the behavior of $P\cO$ and similar products at large $|y_j|$ that 
determines what is happening. For the CLE it is also clear that a very 
large cutoff will practically not be felt, because the system very rarely 
will make contact with it. This is borne out by our numerics which 
clearly shows convergence to the
limit of infinite cutoff. For the FPE, on the other hand
the issue is less clear, because there are very large boundary terms 
arising from the gradients of the drift across the `seam'. In any case, 
for the FPE we cannot directly compare with the cutoff-free results, 
because these do not exist.

\section{Case studies}
\label{secV}

\subsection{The U(1) one-link model}

To understand in more detail how boundary terms affect the behaviour of 
complex Langevin simulations, we studied in some detail the U(1) one-link
model in the hopping (HDM) approximation that was already discussed 
in Ref.~\cite{aartsstam} for $N_I=0$.

The action is
\be
S=-\beta \cos z -\kappa \cos(z-i\mu)= - a\cos(z-ic),
\ee
with 
\be
a=\sqrt{(\beta+\kappa e^\mu)(\beta+\kappa e^{-\mu)}},
\ee
\be
c=\frac{1}{2}\ln\frac{\beta+\kappa e^\mu}{\beta+\kappa e^{-\mu}}.
\ee
The complex drift force is correspondingly
\begin{align}
K=-S'=& -\beta\sin z -\kappa \sin (z-i\mu) \notag \\
=&  -a \sin (z-ic),
\end{align}
and the two components of the drift read
\begin{align}
K_x&=-{\rm Re}\, S'= -a\sin x\cosh(y-c), \\
K_y&=-{\rm Im}\, S'= -a\cos x \sinh(y-c).
\end{align}
As discussed in Ref.\ \cite{aartsstam} there are two fixed points at 
$(x,y)=(0,c)$ and $(x,y)=(\pi,c)$;  the first one is attractive, the second 
one repulsive. 

A special feature of this model is that for $y=c$ the drift is purely in 
$x$ direction. If in addition $N_I=0$, the Langevin process will never 
leave the line $y=c$ if it starts there (we emphasize that the properties 
discussed in this paragraph do not hold for the full 
U(1) one-link model, which was studied in detail in Ref.\ \cite{aartsstam}).
It is therefore straightforward to find an explicit solution to the stationary 
FPE,  
\be
\label{FPEsol}
P(x,y;\infty)\propto\, e^{-S(x+ic)}\delta(y-c). 
\ee
It follows that this model is actually equivalent to one with a real action, 
once we shift $y\to y+c$ and replace $\beta$ by $a$ ($a$ and $c$ now embody 
the dependence on the parameters of the model).
The numerics presented below show that the line $y=c$ is an 
attractor for the Langevin process; this indicates that the solution 
(\ref{FPEsol}) is unique (with proper normalization). 
These properties imply that for $N_I=0$ the dynamics is completely understood.

When $N_I>0$, the presence of the repulsive fixed point is responsible for 
the occurrence 
of large excursions, which are well known to be the scourge of complex 
Langevin simulations. For large $y$ the drift terms dominate over the 
noise, and the Langevin process is essentially just a deterministic motion; 
the `classical' trajectories are given by
\be
z(t)=i\ln\frac{1-iCe^{-a t}}{1+iCe^{-a t}}+ic,
\ee
where the complex integration constant $C$ is related to the starting point 
by
\be
C=\tan\left(\frac{z(0)-ic}{2}\right).
\ee
It is easy to see that all trajectories, except those starting on the 
unstable trajectories (${\rm Im}\, z(0)=\pi$ ) are attracted to the 
stable fixed point at $z=ic$.

\begin{figure}
\begin{center}
\includegraphics[width=0.85\columnwidth]{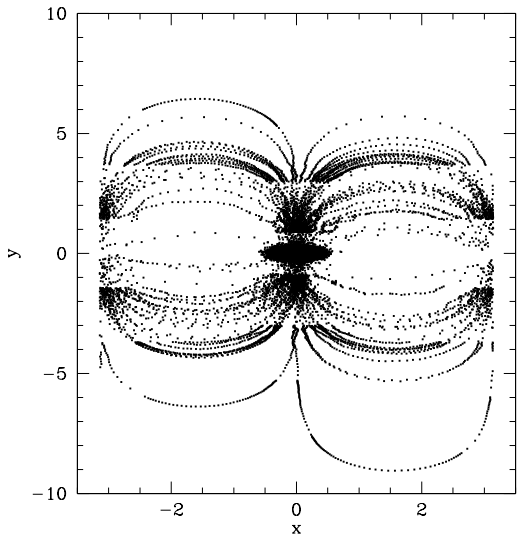}
\caption{Scatter plot for the $U(1)$ one-link model at 
$\beta=1,\\\kappa=0,\  N_I=1$ with reduced noise (see text).
}
\label{u1}
\end{center}
\end{figure}

As an illustration we show in Fig.\ \ref{u1} a scatter plot of a Langevin 
simulation clearly exhibiting the classical orbits. There were 500 update 
steps between consecutive points and we used $N_I=1$. To enhance the 
classical features, this simulation was done with the noise terms in the 
CLE (see Eqs.\ (\ref{eqcle1}, \ref{cledisc}) below) suppressed by a factor 
10; this is of course equivalent to replacing $\epsilon$ by $\epsilon/100$ 
while multiplying the force terms by a factor of 100. This scatter plot 
should capture some generic features that will also be present for other 
choices of the model parameters.

In order to numerically study the role of boundary terms and large $|y|$ 
values at nonzero $N_I$, we introduce a cutoff in imaginary direction, 
placed symmetrically around $y=c$, i.e.\ $-Y+c<y<Y+c$, and impose periodic 
 boundary conditions.
To see the effect of this cutoff, we compute numerically the expectation 
values of the observables $\exp(ikz)$ for $k=\pm1,\pm2$, for various values 
of $N_I$ and $Y$, both by simulating the Langevin process and by 
numerical solution of the FPE. Note that these observables grow exponentially 
at large $y$.
The exact values are given by
\be
\bra e^{ikx}\ket=\frac{I_k(a)}{I_0(a)} e^{-kc},
\ee
where $I_k(a)$ are the modified Bessel functions of the first kind.

The Langevin process is discretized in the usual way, 
\begin{align}
\label{eqcle1}
x_{n+1}=x_n+\epsilon K_x(x_n,y_n)+\sqrt{\epsilon N_R}\,\eta_{x,n}\\
y_{n+1}=y_n+\epsilon K_y(x_n,y_n)+\sqrt{\epsilon N_I}\,\eta_{y,n}
\label{cledisc}
\end{align}
where $\eta_{x,n}$ and $\eta_{y,n}$ are pseudorandom numbers with zero mean and variance 2. 
We use periodic boundary conditions in the imaginary direction, as stated 
above. We also use an adaptive step size \cite{Aarts:2009dg}, choosing 
$\epsilon$ such that the product of $\epsilon |K_x+iK_y|\le 0.001$.
To estimate the statistical error we run 100 trajectories with 
independent random starting points.

\begin{figure}[t]  
\includegraphics[width=0.9\columnwidth]{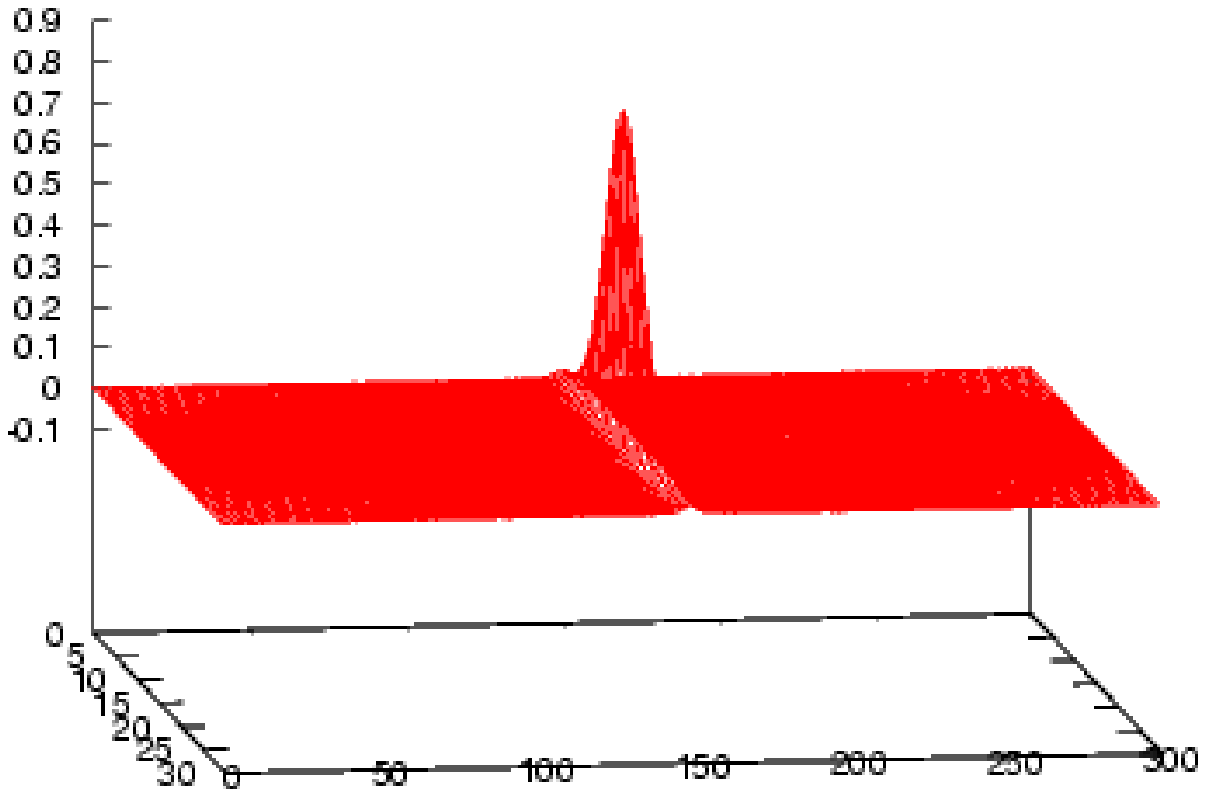}
\vspace*{0.2cm}
\includegraphics[width=0.9\columnwidth]{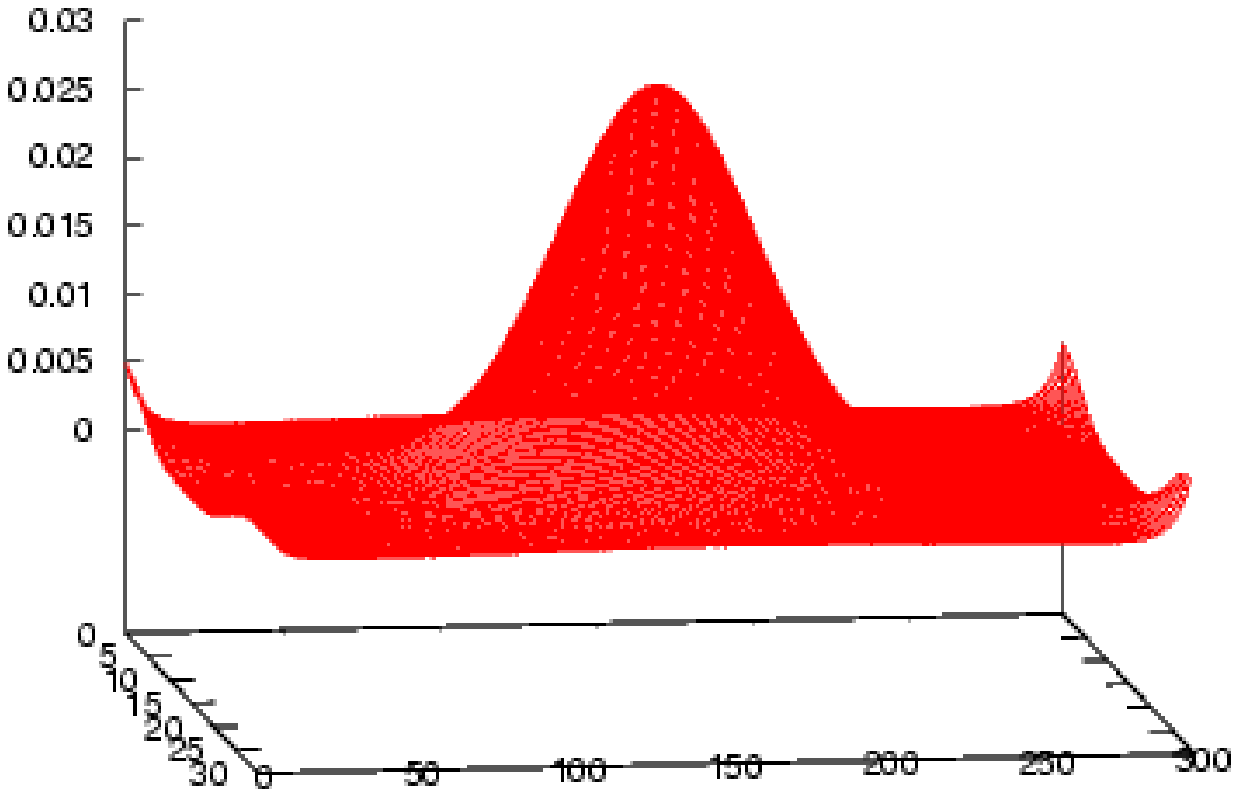}
\vspace*{0.2cm}
\includegraphics[width=0.9\columnwidth]{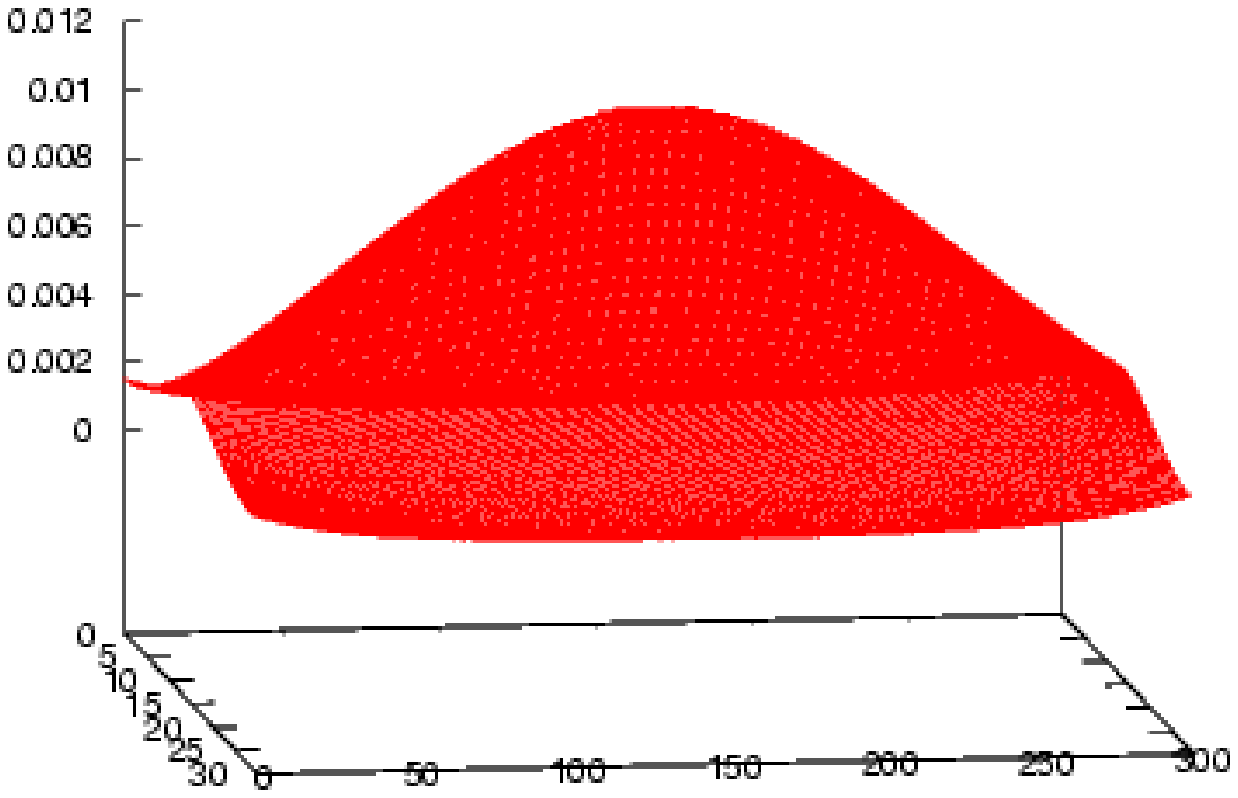}
 \caption{Distributions $P(x,y;t\to\infty)$ in the U(1) one-link model, 
obtained from a numerical solution of the real FPE, for various values of 
$N_I$: $N_I=0.0001$ (top), $0.01$ (middle), $0.1$ (bottom). See the main 
text for further details.
 }
\label{f.P_dis}
\end{figure}

\begin{figure}[t]
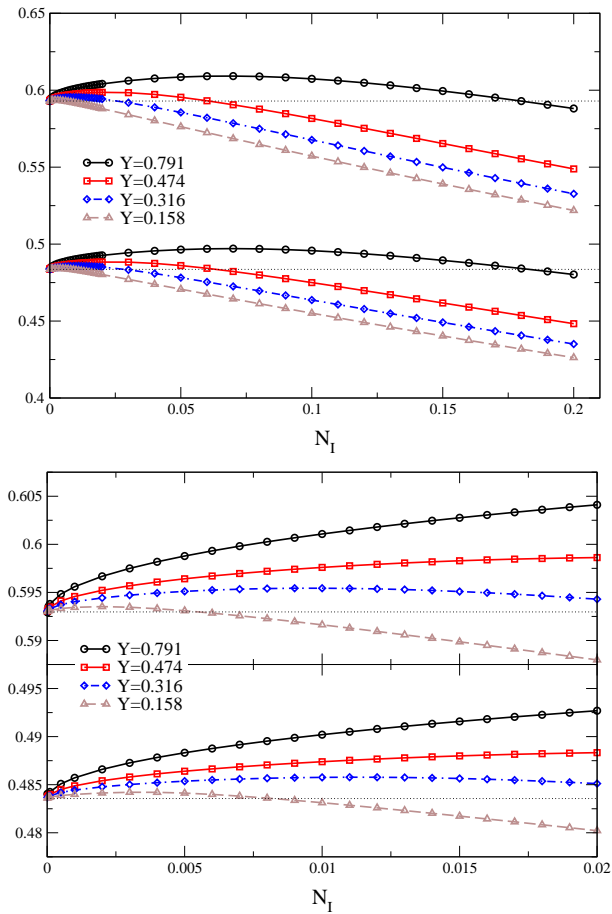
  
\begin{center}
\includegraphics[width=.9\columnwidth]{fig_U1_NI.eps} \\
\;\;\includegraphics[width=.93\columnwidth]{fig_U1_NIs2.eps}
\end{center}
\caption{$N_I$ dependence of $\mbox{Re}\,\bra e^{iz}\ket$ (lower points) 
and  $\mbox{Re}\,\bra e^{-iz}\ket$ (higher points) from FPE 
for various values of 
the cutoff $Y$. The bottom figure zooms in on smaller values of $N_I$. 
The lines are guides to the eye, the
horizontal dotted lines indicate the correct results (0.483564 and 
0.592966 respectively).
  }
\label{f.ni-dep}
\end{figure}  

To solve the FPE  numerically, we employ 
the periodicity in $x$ and consider the Fourier decompositions
\begin{align}
\widehat \rho(k;t)= & \int \frac{dx}{2\pi}\, e^{ikx}\rho(x;t), \\
\widehat P(k,y;t)=& \int \frac{dx}{2\pi}\, e^{ikx} P(x,y;t),
\end{align}
with the inverse transformations given by
\begin{align}
\rho(x;t)=& \sum_{k=-\infty}^\infty e^{-ikx} \widehat \rho(k;t), \\
P(x,y;t)=&\sum_{k=-\infty}^\infty e^{-ikx} \widehat P(k,y;t).
\end{align}
The  FPE can be rewritten in terms of these modes as
\begin{align}
&\partial_t \widehat P(k,y;t)=(-N_R k^2 +N_I\partial_y^2)\widehat P(k,y;t) 
\cr
&+\,\frac{a}{2} \cosh(y-c)\Big[(k+1)\widehat P(k-1,y;t)
\notag \\
& \hspace*{2.2cm} -
(k-1)\widehat P(k+1,y;t)\Big]\notag\\
&+\,\frac{a}{2}\sinh(y-c)\partial_y\left[\widehat P(k-1,y;t)+
\widehat P(k+1,y;t)\right]. \;\;\;\;
\end{align}
 This equation is solved numerically with a discretized time step 
$\epsilon=10^{-5}$ and a spatial discretization in $y$ of 
$\delta=\sqrt{\epsilon}$. We vary the cutoff $Y$ in the $y$ direction from 
$10\delta$ up to $500\delta$ ($1500\delta$ in some cases). 
In the $x$ direction, with $-\pi<x\leq \pi$, we use $20-30$ points.  
We found that convergence was reached after $10^6$ time 
steps, corresponding to a Langevin time $t \sim 10$. We found convergence 
for all the values of $N_I$ and $Y$ studied. For $N_I=0$, we were not able 
to solve the FPE numerically, due to the singular behaviour, but the 
solution is known analytically, see Eq.\ (\ref{FPEsol}).

We fix the parameters of the model to be 
\be
\beta=1.0,\;\;\;\; \kappa=0.25,\;\;\;\; \mu=0.5.
\ee
 In Fig.\ \ref{f.P_dis} we show some examples of the real probability 
distribution $P(x,y,t)$ at large Langevin time $t \sim 20$ for various 
$N_I$ and a given cutoff $Y=0.474$. The $y$ direction goes from 
left to right and the compact $x$ direction from back to front. We observe 
that at small $N_I= 0.0001$, the distribution resembles the analytic 
result at $N_I=0$. The distribution is very narrow in the $y$ direction 
and boundary effects are not expected to play a role. Increasing $N_I$ 
results in a wider distribution, and boundary effects become clearly 
visible. The apparent non-smooth behaviour at the edge $|y-c|=Y$ is to be 
expected from the discontinuity across the seam, discussed at the end of 
the previous section.

\begin{figure}[t]  
\begin{center}
\includegraphics[width=0.9\columnwidth]{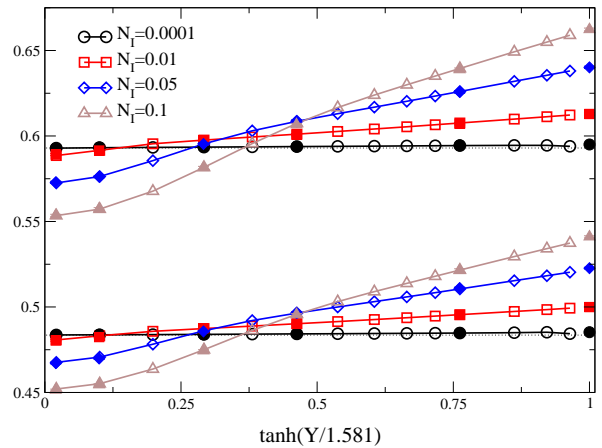}
\end{center}
 \caption{Cutoff ($Y$) dependence of $\mbox{Re}\,\bra e^{iz}\ket$ (lower 
data) and $\mbox{Re}\,\bra e^{-iz}\ket$ (upper data) for various values 
of $N_I$, for FPE (open symbols) and CLE (full symbols, note that the 
errorbars are much smaller that the points).
The lines are guides to the eye. 
The horizontal dotted lines indicate the correct results.
 }
\label{f.L-dep}
\end{figure}

After obtaining the distribution, expectation values of observables follow 
from Eq.\ (\ref{eq:OP}). In Tables 1 -- 7 (see end of paper)  we compare 
the results of the Langevin simulation and the FPE for increasing values 
of $N_I$ from $0$ up to $N_I=0.1$. Note that the Langevin equation can be 
solved without a cutoff ($Y=\infty$). The imaginary parts of the 
observables are consistent with zero. The data of the tables are also 
summarized in Figs.\ \ref{f.ni-dep}, \ref{f.L-dep} (the FPE and the CLE 
data are indistinguishable at the scale of the figures; the rightmost 
points in Fig.\ \ref{f.L-dep} correspond to $Y=4.74$ for FPE and $\infty$ 
for CLE).

The following facts can be inferred from these results:\\
(1) CLE and FPE give rather similar results, but sometimes they differ by 
several $\sigma$ (statistical error of the CLE simulation).\\ 
(2) All the data show a clear dependence on $N_I$, in contrast to the 
conclusion of the formal arguments. For larger $N_I$ values both CLE and 
FPE give results different from the exact values.\\
(3) The best results are generally obtained for the smallest $N_I$. In 
this case there is also the weakest $Y$ dependence, in fact no $Y$ 
dependence whatsoever for $N_I=0$.\\

Obviously the presence of the cutoff and periodic boundary conditions 
affects the CLE and FPE in a similar way. But the 
$N_I$ dependence shows the failure of the formal argument, even for 
$Y=\infty$. At least for $k=\pm1$ one has a clear case of `convergence to 
a wrong limit'.
The data of the CLE with $Y=\infty$ and $k=\pm 2$ are actually not really 
converged, except for $N_I=0.$ Observing the scatter plots for different 
Langevin times it appears unclear whether the observables $\exp(\pm 
2iz)$  really reach an equilibrium distribution -- they tend to drift 
out to infinity, whereas their averages, while remaining small, suffer 
from huge fluctuations. On the other hand the observables $\exp(\pm iz)$ 
seem to reach a stable distribution for $Y=\infty$ and any $N_I$.

\subsection{The model of Guralnik and Pehlevan}

Guralnik and Pehlevan \cite{guralnik} studied an instructive toy model on 
$\cM=\R$ and $\cM_c=\C$, called GP model henceforth. Its action is
\be 
\label{GPaction}
S = -i\beta\left(z+\frac{1}{3}z^3\right),
\ee
and was studied in connection with PT invariant but non-hermitian 
Hamiltionians (where PT indicates the combined action of parity and time 
reversal) \cite{bender}.

\begin{figure}[t]
\begin{center}
\includegraphics[width=0.85\columnwidth]{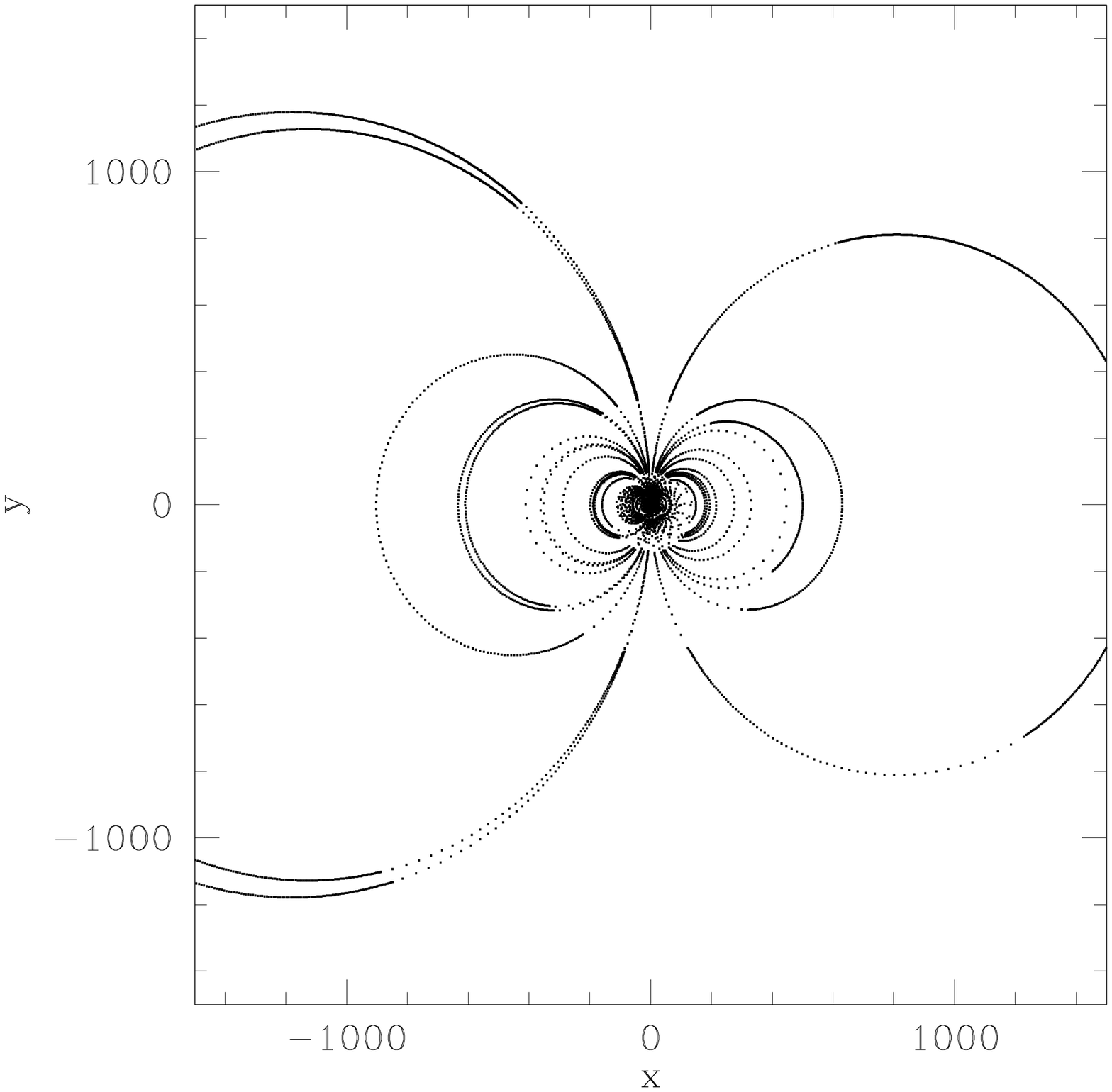}
\caption{Scatter plot in the GP model with $\beta=1$, $N_I=1$.}
\label{guralnikclass}
\end{center}
\includegraphics[width=0.85\columnwidth]{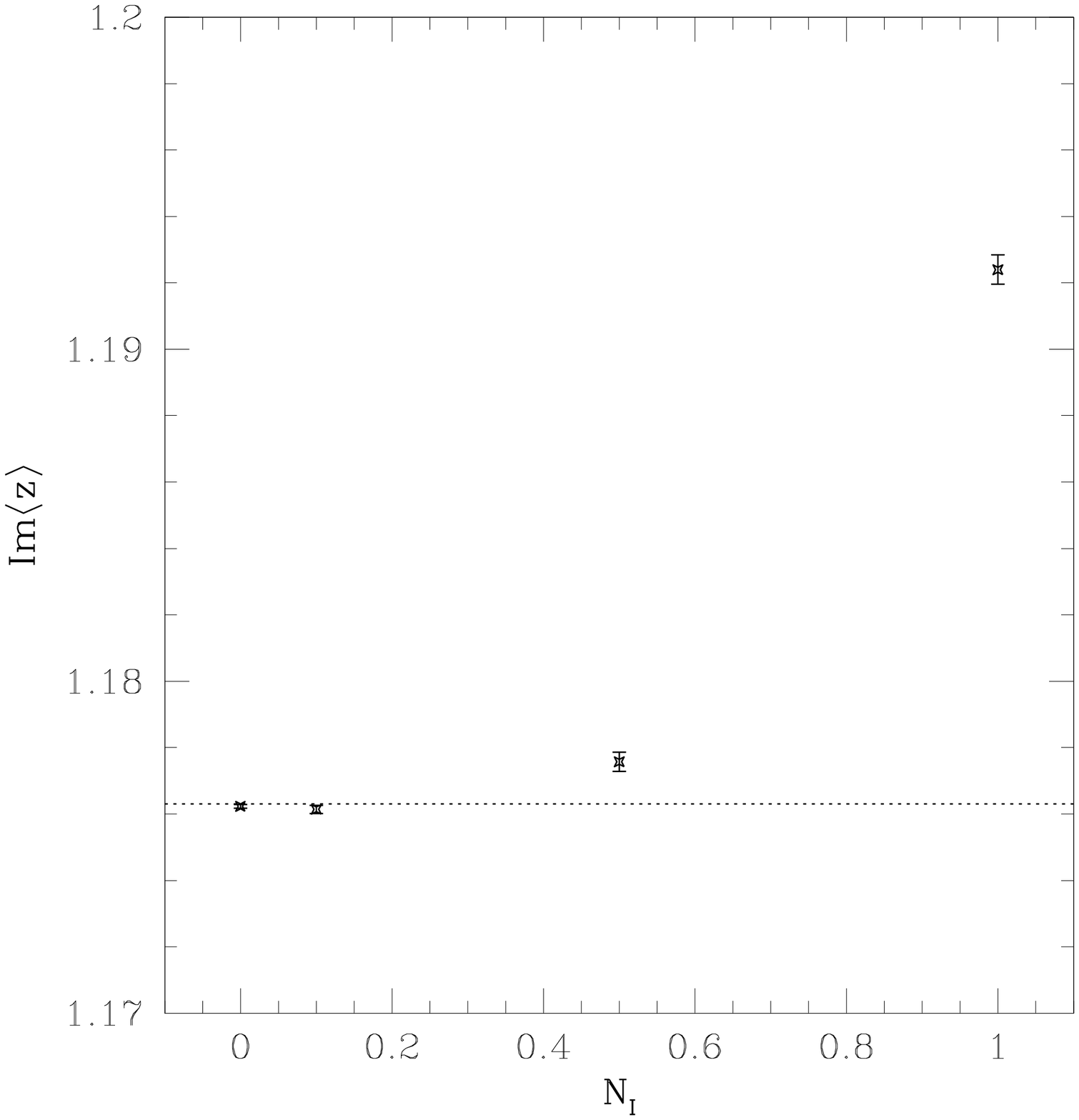}
\caption{$N_I$ dependence of ${\rm Im}\,\bra z\ket$ in the GP model at 
$\beta=1$.}
\label{Imz} 
\end{figure}

The action (\ref{GPaction}) leads to the drift forces
\be
K_x=-2\beta x y, \;\;\;\; K_y=\beta(1+x^2-y^2).
\ee
There is a stable fixed point at $z=i$ and an unstable one at $z=-i$. The 
`classical trajectories' obtained by leaving out the noise are given by
\be
z(t)=\frac{z_0+i\tanh(\beta t)}{1-iz_0\tanh(\beta t)}.
\ee
 Since this is a M\"obius transformation from $w=\tanh(\beta t)$ to $z(t)$ 
the trajectories are circles. They can be imagined to emerge from the 
unstable fixed point $z=-i$ at $t=-\infty$ and go to the stable fixed 
point $z=i$ as $t\to\infty$.
 Those classical trajectories again can be seen clearly in the large 
excursions, since there the noise becomes negligible. Fig.\ 
\ref{guralnikclass} shows the result from a Langevin simulation at 
$N_I=1.0$ and $\beta=1.0$ (in this case there are 50000 update steps 
between two consecutive points).

\begin{figure}[t]  
\begin{center}
\includegraphics[width=0.85\columnwidth]{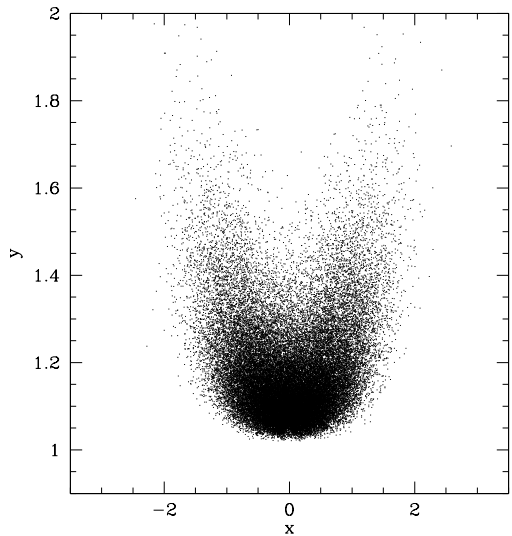}
\caption{Scatter plot in the GP model for $\beta=1, N_I=0$.}
\label{plots2}
\end{center}
\end{figure}

Guralnik and Pehlevan give the exact results of the first three moments at  
$\beta=1$, they are
\begin{align}
&\bra z \ket = -i\frac{{\rm Ai'}(1)}{{\rm Ai}(1)}=1.1763i,\notag\\
&\bra z^2\ket =-1,\notag\\ 
&\bra z^3\ket = i-\bra z \ket=-0.1763 i. 
\end{align}
They solved the discretized Langevin equation numerically, using $N_I=0.001$, 
and obtained good agreement with those exact results.
We did some more and probably longer simulations at $N_I=0$, $0.1$, $0.5$, 
$1.0$. The results for ${\rm Im}\, \bra z \ket$ are shown in 
Fig.\ \ref{Imz};  again they show a clear dependence on $N_I$, in 
conflict with the formal reasoning. While for small $N_I$ there is 
agreement with the exact result, for larger $N_I$ we again have 
convergence to the wrong limit.

\begin{figure}[t]  
\includegraphics[width=0.85\columnwidth]{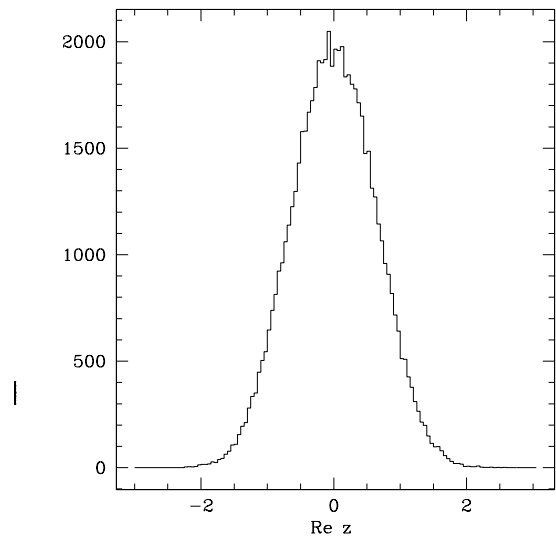}
\includegraphics[width=0.85\columnwidth]{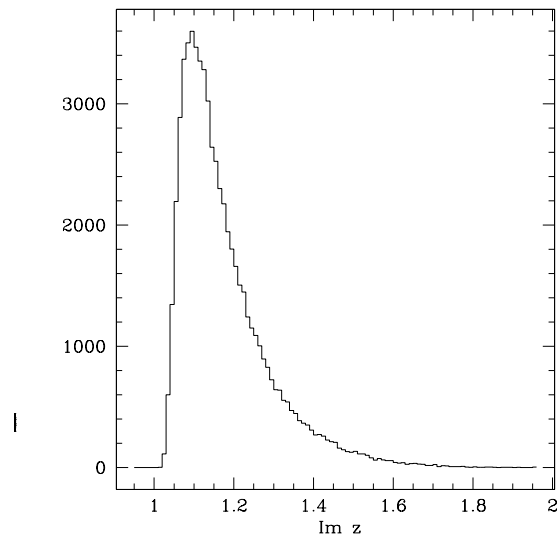}
\caption{Distribution of Re $z$ (top) and Im $z$ (bottom) in the GP model with 
$\beta=1$, $N_I=0$.}
\label{plots}
\end{figure}

The results for ${\rm Re}\,\bra z^2 \ket$ show a similar behaviour, except 
that at $N_I=1.0$ the result is $-0.951(44)$, with a statistical error that 
is huge compared to the error $0.0007$ found at $N_I=0.5$. 
 The data for ${\rm Im }\,\bra z^3 \ket$ show an even more dramatic failure
for larger $N_I$: they diverge for $N_I=1.0$, whereas for $N_I=0.5$ 
fluctuations are large. 
 Finally we measured ${\rm Im}\,\bra e^{iz} \ket$, which has an exact value 
of $2.2624$. Here divergence becomes manifest already for $N_I=0.5$ (but 
it might occur for all $N_I>0$). 

In order to have an idea of the equilibrium distribution for $N_I=0$ we 
show in Fig.\ \ref{plots2} a scatter plot of 50000 configurations in the 
complex plane. Non-Gaussian behaviour is quite clear from this plot. 
Noticeable is the appearance of sharp edges of the distribution, possibly 
indicating jumps, but no $\delta$ functions like in the $U(1)$ case in the 
hopping expansion. To obtain these pictures we sampled over 60000 points 
taken at equal intervals of 0.5 in Langevin time. Similar distributions 
have been observed in Ref.\ \cite{aartsstam} for the full U(1) one-link 
model. The non-Gaussian character is further demonstrated in Fig.\ 
\ref{plots}: the histograms for both ${\rm Re}\, z$ and ${\rm Im}\, z$ 
deviate strongly from a Gaussian distribution.

\section{Tentative conclusions}
\label{secVI} 
 
\subsection{Explanation of wrong results}

We will now try to interpret these findings. They are quite analogous
in the two model cases and we will try to reach some conclusions that can 
be generalized to more realistic models.

The question of convergence vs. divergence apparently depends on the values 
of $N_I$, but it is more plausible that there is no qualitative difference 
between the different positive values of $N_I$, only the time needed to 
observe the asymptotic behaviour is different. On the other hand for 
$N_I=0$ there seems to be really a qualitative difference: the 
distributions develop discontinuities or even $\delta$ functions; but more 
important is the fact that they seem to drop very rapidly in the imaginary 
direction.

We tentatively conclude that for $N_I=0$ the systems relax to equilibrium 
measures that show at least exponential decay in imaginary direction
(in our simple $U(1)$ model this decay is of course much stronger -- the 
measure is zero for $|y|>c$).

For $N_I>0$ the situation is less clear, but it seems that in the $U(1)$ 
model we get a decay at least like $\exp(-|y|)$, but probably not stronger 
than any exponential.  In the model of Guralnik and Pehlevan the data 
suggest a power-like decay (the power appears to be near~2).

To sum up our tentative conclusions:

{\it There is a unique equilibrium distribution $P(x,y;\infty)$ for the
Langevin processes for any $N_I\ge 0$, but for $N_I>0$ it shows limited 
decay, especially in the imaginary direction.}

The type of decay for $N_I>0$ depends on the model considered; but for
some observables their growth in imaginary direction may conspire with 
the falloff of the equilibrium measure in such a way that we obtain 
convergence to a wrong limit. In any case one should not expect 
convergence of the mean value for {\it all} holomorphic observables.
 
For very small $N_I$ and limited simulation time, the behaviour is of course 
indistinguishable from the one at $N_I=0$, and one may reach a 
quasi-convergence to the right limit, even if an infinitely long 
simulation would diverge. 

If we try to generalize boldly from our toy $U(1)$ model to real lattice 
gauge theories, we expect that for $N_I>0$ and for most interesting 
observables, such as Wilson loops, Polyakov loops etc., we have to expect 
boundary terms contributing even as the boundary is sent to infinity. For 
multiply charged loops and $N_I>0$ the situation could be even worse: 
those boundary contributions may diverge as the boundary is moved to 
infinity. But these problems probably will not occur for $N_I=0$ 
and do not show up at very small values of $N_I$ either, at least if 
simulations are not run excessively long.

We realize that definite conclusions about the falloff of the probability density $P(x,y)$ in the 
$y$ direction have not yet been reached; we intend to return to a more detailed study of this 
question in a future paper.

\subsection{Practical conclusions}

The conclusions for practical applications are quite straightforward:

(1) In complex Langevin simulations one should use $N_I=0$. If one wants 
to do a random walk simulation or an iterate of the Fokker-Planck 
operator, this is not possible, but one should make sure that $N_I \ll 1$.

(2) Always validate the simulations by comparison with other trusted 
computations for parameter values where other methods are available.  

(3) Wilson or Polyakov loops of higher charge will generally have much 
larger fluctuations. In general they are not needed for physics 
applications, but they may be worth looking at because they can give 
information about the probability distribution. 

\vspace*{0.3cm}

\acknowledgments

We thank Frank James and Denes Sexty for discussion. 
I.-O.S.\ thanks the MPI for Physics M\"unchen and Swansea University for 
hospitality. G.A.\ is supported by STFC.

\onecolumngrid

\begin{center}

\begin{table}[b]
\begin{tabular}{|c|c||l|l|l|l|}
 \hline
$Y$   &   &\ \ \ \ \    $\bra U\ket $ \ \ \ \ \ &\ \ \ \ \  $\bra U^{-1}\ket$\ \ 
\ \ \ & \ \ \ \ \ $\bra U^2\ket$\ \ \ \ \ &\ \ \ \ \  $\bra U^{-2}\ket$\ \ \ \ \   \\ 
 \hline
 \hline

  0.032 & CLE  &  0.48331(047)   &  0.59265(058)   &  0.12311(037)   &  
0.19865(055) \\         
 \hline
 \hline
 0.158& CLE  &  0.48331(047)   &  0.59265(058)   &  0.13211(037)   &  
0.19865(055) \\ 
  \hline
\hline 
 0.474 & CLE  &  0.48331(047)   &  0.59265(058)   &  0.13211(037)   &  
0.19865(055) \\ 
 \hline
 \hline
 0.790 & CLE  &  0.48331(047)   &  0.59265(058)   &  0.13211(037)   &  
0.19865(055) \\ 
 \hline
 \hline
1.581& CLE  &  0.48331(047)   &  0.59265(058)   &  0.13211(037)   &  
0.19865(055)    \\ 
 \hline
 \hline
$\infty$ &CLE&  0.48331(047)   &  0.59265(058)   &  0.13211(037)   &  
0.19865(055)   \\ 
 \hline 
 \hline 
exact &   &  0.48356  &  0.59297   &  0.13065  &  0.19646 \\         
 \hline

\end{tabular} 
\caption{$N_I=0.0$}

\vspace*{1cm}

\begin{tabular}{|c|c||l|l|l|l|}
 \hline
$Y$   &   &\ \ \ \ \    $\bra U\ket $ \ \ \ \ \ &\ \ \ \ \  $\bra U^{-1}\ket$\ \ \ \ \ & \ \ \ \ \ 
$\bra U^2\ket$\ \ \ \ \ &\ \ \ \ \  $\bra U^{-2}\ket$\ \ \ \ \   \\ 
 \hline
 \hline

  0.032  & CLE  &  0.48344(046)   &  0.59281(057)   &  0.13154(036)   &  
0.19778(054) \\          
   & FPE &  0.48358   &  0.59297   &  0.13065   &  0.19646 \\         
 \hline
 \hline
 0.158& CLE  &  0.48369(047)   &  0.59313(057)   &  0.13194(036)   &  
0.19835(055) \\ 
   & FPE &  0.48361   &  0.59302   &  0.13065      &  0.19645 \\         
 \hline
 \hline
 0.474 & CLE  &  0.48370(047)   &  0.59313(058)   &  0.13188(036)   &  
0.19822(054) \\ 
   & FPE &  0.48376   &  0.59307   &  0.13066   &  0.19639 \\         
 \hline
 \hline
  0.790 & CLE  &  0.48372(048)   &  0.59319(059)   &  0.13186(036)   &  
0.19812(055)  \\ 
   & FPE &  0.48405   &  0.59292   &  0.13076   &  0.19618 \\         
 \hline
 \hline
1.581& CLE  &  0.48382(047)   &  0.59330(058)   &  0.13183(037)   &  
0.19792(056)  \\ 
   & FPE &  0.48647   &  0.59033   &  0.13318   &  0.19449 \\         
 \hline
 \hline
$\infty$ &CLE&  0.48396(047)   &  0.59345(058)   &  0.13089(042)   &  
0.19685(056)    \\ 
\hline 
 \hline
exact &   &  0.48356  &  0.59297   &  0.13065  &  0.19646 \\         
 \hline
 
\end{tabular} 
\caption{$N_I=0.00001$}
\end{table}


\begin{table}
\begin{tabular}{|c|c||l|l|l|l|}
 \hline
$Y$   &   &\ \ \ \ \    $\bra U\ket $ \ \ \ \ \ &\ \ \ \ \  $\bra U^{-1}\ket$\ \ \ \ \ & \ \ \ \ \ 
$\bra U^2\ket$\ \ \ \ \ &\ \ \ \ \  $\bra U^{-2}\ket$\ \ \ \ \   \\ 
 \hline
 \hline

  0.032  & CLE  &  0.48345(050)   &  0.59280(062)   &  0.13114(035)   &  
0.19714(052) \\          
   & FPE &  0.48366   &  0.59290   &   0.13066   &  0.19646 \\         
 \hline
 \hline 
  0.158& CLE  &  0.48385(054)   &  0.59335(067)   &  0.13153(039)   &  
0.19771(059) \\ 
   & FPE &  0.48371   &  0.59313   &  0.13064  &  0.19644 \\         
 \hline
 \hline 
  0.474 & CLE  &  0.48402(052)   &  0.59357(064)   &  0.13148(037)   &  
0.19762(056) \\ 
   & FPE &  0.48399   &  0.59348  &  0.13059   &  0.19637 \\         
 \hline
 \hline 
 0.790 & CLE  &  0.48425(053)   &  0.59377(065)   &  0.13146(038)   &  
0.19764(057) \\ 
   & FPE &  0.48425   &  0.59379   &  0.13056   &  0.19627 \\         
 \hline
 \hline 
1.581 & CLE  &  0.48485(051)   &  0.59454(063)   &  0.13115(036)   &  
0.19725(054)  \\ 
   & FPE &  0.48476   &  0.59437   &  0.12363   &  0.19583 \\         
 \hline
 \hline 
$\infty$ &CLE&  0.48527(051)   &  0.59504(063)   &  0.12944(047)   &  
0.19387(079)   \\ 
 \hline 
 \hline
exact &   &  0.48356  &  0.59297   &  0.13065  &  0.19646 \\         
 \hline
 
\end{tabular} 
\caption{$N_I=0.0001$}

\vspace*{1cm}

\begin{tabular}{|c|c||l|l|l|l|}
 \hline
$Y$   &   &\ \ \ \ \    $\bra U\ket $ \ \ \ \ \ &\ \ \ \ \  $\bra U^{-1}\ket$\ \ \ \ \ & \ \ \ \ \ 
$\bra U^2\ket$\ \ \ \ \ &\ \ \ \ \  $\bra U^{-2}\ket$\ \ \ \ \   \\ 
 \hline
 \hline
  0.032  & CLE  &  0.48342(049)   &  0.59293(060)   &  0.13095(038)   &  
0.19692(057) \\          
   & FPE &  0.48373   &  0.59226   &  0.13074   &  0.19602 \\         
 \hline
 \hline 
  0.158& CLE  &  0.48410(051)   &  0.59346(062)   &  0.13068(037)   &  
0.19632(056) \\ 
   & FPE & 0.48399   &  0.59344   &  0.13065   &  0.19647 \\         
 \hline 
 \hline 
   0.474 & CLE  &  0.48538(051)   &  0.59497(064)   &  0.13090(038)	&  
0.19702(058) \\ 
   & FPE &  0.48488   &  0.59457   &  0.13050   &  0.19623 \\         
 \hline
 \hline 
 0.790 & CLE  &  0.48619(051)   &  0.59605(061)   &  0.13066(041)   &  
0.19655(063)   \\ 
   & FPE &  0.48572   &  0.59559   &  0.13027   &  0.19589 \\         
 \hline
 \hline 
1.581 & CLE  &  0.48762(052)   &  0.59761(062)   &  0.12934(044)   &  
0.19503(064)    \\ 
   & FPE &  0.48729   &  0.59753   &  0.12934   &  0.19449 \\         
 \hline
 \hline 
$\infty$ &CLE&  0.48890(050)   &  0.59942(063)   &  0.12392(068)   &  
0.18602(086)  \\ 
 \hline 
 \hline
exact &   &  0.48356  &  0.59297   &  0.13065  &  0.19646 \\         
 \hline
 
\end{tabular} 
\caption{$N_I=0.001$}

\vspace*{1cm}

\begin{tabular}{|c|c||l|l|l|l|}
 \hline
$Y$   &   &\ \ \ \ \    $\bra U\ket $ \ \ \ \ \ &\ \ \ \ \  $\bra U^{-1}\ket$\ \ \ \ \ & \ \ \ \ \ 
$\bra U^2 \ket$\ \ \ \ \ &\ \ \ \ \  $\bra U^{-2}\ket$\ \ \ \ \   \\ 
 \hline
 \hline

  0.032  & CLE  &  0.48047(049)   &  0.58917(060)   &  0.12935(035)   &  
0.19450(053) \\          
   & FPE &  0.48073   &  0.58856   &  0.12894   &  0.19328 \\         
 \hline
 \hline 
  0.158& CLE  &  0.48240(051)   &  0.59150(063)   &  0.13039(037)   &  
0.19613(056) \\ 
   & FPE &  0.48313   &  0.59163  &  0.13055   &  0.19584 \\         
 \hline
 \hline 
  0.474 & CLE  &  0.48747(053)   &  0.59753(067)   &  0.13080(041)   &  
0.19645(059) \\ 
   & FPE &  0.48739   &  0.59760   &  0.13064   &  0.19646 \\         
 \hline
 \hline 
  0.790 & CLE  &  0.49012(055)   &  0.60078(069)   &  0.12952(045)   &  
0.19437(069)  \\ 
   & FPE &  0.49020   &  0.60107   &  0.12984   &  0.19525 \\         
 \hline
 \hline 
1.581 & CLE  &  0.49532(052)   &  0.60714(068)   &  0.12647(054)   &  
0.19056(085)  \\ 
   & FPE &  0.49550   &  0.60758   &  0.12668  &  0.19051 \\         
 \hline
 \hline 
$\infty$ &CLE&  0.49995(049)   &  0.61287(063)   &  0.10733(202)   &  
0.16132(376) \\ 
 \hline 
 \hline
exact &   &  0.48356  &  0.59297   &  0.13065  &  0.19646 \\         
 \hline
 
\end{tabular} 
\caption{$N_I=0.01$}
\end{table}


\begin{table}
\begin{tabular}{|c|c||l|l|l|l|}
 \hline
$Y$   &   &\ \ \ \ \    $\bra U\ket $ \ \ \ \ \ &\ \ \ \ \  $\bra U^{-1}\ket$\ \ \ \ \ & \ \ \ \ \ 
$\bra U^2\ket$\ \ \ \ \ &\ \ \ \ \  $\bra U^{-2}\ket$\ \ \ \ \   \\ 
 \hline
 \hline

  0.032  & CLE  &  0.46720(051)   &  0.57290(063)   &  0.12150(034)   &  
0.18269(051) \\          
   & FPE &  0.46763   &  0.57253    &  0.12118   &  0.18165 \\         
 \hline 
 \hline
  0.158& CLE  &  0.46985(054)   &  0.57620(066)   &  0.12335(037)   &  
0.18557(055) \\ 
   & FPE &  0.47070   &  0.57629   &  0.12363   &  0.18533 \\         
 \hline 
 \hline
  0.474 & CLE  &  0.48570(057)   &  0.59533(071)   &  0.13167(042)   &  
0.19766(065) \\ 
   & FPE &  0.48608  &  0.59542   &  0.13184   &  0.19793 \\         
 \hline 
 \hline
  0.790 & CLE  &  0.49664(055)   &  0.60871(068)   &  0.13166(046)   &  
0.19731(074)  \\ 
   & FPE &  0.49644   &  0.60858   &  0.13166   &  0.19799 \\         
 \hline 
 \hline
1.581 & CLE  &  0.51104(056)   &  0.62580(080)   &  0.12366(074)   &  
0.18370(114) \\ 
   & FPE &  0.51055   &  0.62601   &  0.12363   &  0.18595 \\         
 \hline 
 \hline
$\infty$ &CLE&  0.52272(051)   &  0.64013(069)   &  0.07553(204)   &  
0.11359(334)     \\ 
 \hline 
 \hline
exact &   &  0.48356  &  0.59297   &  0.13065  &  0.19646 \\         
 \hline
 
\end{tabular} 
\caption{$N_I=0.05$}

\vspace*{1cm}

\begin{tabular}{|c|c||l|l|l|l|}
 \hline
$Y$   &   & \ \ \ \ \  $\bra U\ket$ \ \ \ \  \  &\ \ \ \  \   $\bra U^{-1}\ket$\ \ \ \ \ & \ \ \ \ 
\ $\bra U^2\ket$\ \ \ \ \ &\ \ \ \ \  $\bra U^{-2}\ket$\ \ \ \ \   \\ 
 \hline
 \hline

 0.032  & CLE  &  0.45144(051)   &  0.55358(062)   &  0.11267(035)   &  
0.16941(053) \\          
   & FPE &  0.45203   &  0.55343   &  0.11236   &  0.16843 \\         
 \hline 
 \hline
  0.158& CLE  &  0.45456(055)   &  0.55742(067)   &  0.11453(038)   &  
0.17224(054) \\ 
   & FPE &  0.45513   &  0.55722   &  0.11479   &  0.17207 \\         
 \hline 
 \hline 
  0.474 & CLE  &  0.47443(057)   &  0.58151(071)   &  0.12738(044)   &  
0.19117(065) \\   
   & FPE &  0.47501   &  0.58165   &  0.12759  &  0.19136\\         
 \hline 
 \hline
 0.790 & CLE  &  0.49560(053)   &  0.60668(073)   &  0.13448(052)   &  
0.20120(081)   \\ 
   & FPE &  0.49575   &  0.60740  &  0.13446   &  0.20198 \\         
 \hline 
 \hline
1.581& CLE  &  0.52141(093)   &  0.63883(093)   &  0.12345(098)   &  
0.18569(144)  \\ 
   & FPE &  0.52157  &  0.63948   &  0.12426   &  0.18692 \\         
 \hline 
 \hline
$\infty$ &CLE&  0.54104(059)   &  0.66230(082)   &  0.05049(354)   &  
0.07723(522)   \\ 
 \hline 
 \hline
exact &   &  0.48356  &  0.59297   &  0.13065  &  0.19646 \\         
 \hline
 
\end{tabular} 
\caption{$N_I=0.1$}
\end{table}

\end{center}

\end{document}